\documentclass[12pt]{iopart}

\usepackage{graphicx}
\usepackage{amssymb}

\begin{document}
\title{Crossing the phantom divide in an interacting generalized Chaplygin gas}
\author{H. Garc\'{\i}a-Compe\'an\footnote{compean@fis.cinvestav.mx}}
\address{Centro de Investigaci\'on y de Estudios Avanzados del IPN, Unidad
Monterrey, Autopista al Aeropuerto km 9.5, CP. 66600, Apodaca, NL, M\'exico\footnote{On leave of Departamento de F\'{\i}sica, Centro de Investigaci\'{o}n y de Estudios
Avanzados del IPN, M\'{e}xico, A.P. 14-740, 07000 M\'{e}xico D.F., M\'{e}xico}}
\author{G. Garc\'{\i}a-Jim\'enez\footnote{ggarcia@fcfm.buap.mx}}
\address{Facultad de Ciencias F\'{\i}sico Matem\'{a}ticas, Universidad Aut\'{o}noma de
Puebla, P.O. Box 1364, 72000 Puebla, M\'{e}xico}
\author{O. Obreg\'on\footnote{octavio@fisica.ugto.mx}}
\address{Instituto de F\'{\i}sica de la Universidad de Guanajuato, P.O. Box E-143,
37150 Le\'{o}n Gto., M\'{e}xico}
\author{C. Ram\'{\i}rez\footnote{cramirez@fcfm.buap.mx}}
\address{Facultad de Ciencias F\'{\i}sico Matem\'{a}ticas, Universidad Aut\'{o}noma de
Puebla, P.O. Box 1364, 72000 Puebla, M\'{e}xico}
\date{\today}

\begin{abstract}
Unified generalized Chaplygin gas models assuming an interaction between dark energy and dark matter fluids have been previously
proposed. Following these ideas, we consider a particular relation between dark densities, which allows the
possibility of a time varying equation of state for dark energy that crosses the
phantom divide at a recent epoch. Moreover, these densities decay
during all the evolution of the Universe, avoiding a Big Rip. We
find also a scaling solution, i.e. these densities are
asymptotically proportional in the future, which contributes to the
solution of the coincidence problem.

\end{abstract}

\maketitle

\section{Introduction}

Today we confront the challenge in cosmology of explaining the
observation that, in the recent past, our Universe has experimented
a phase of accelerated expansion \cite{Riess}. This acceleration is
attributed to the mysterious dark energy (for a review see
\cite{Peebles}), whose understanding is nowadays a major problem.
Although the simplest candidate to play the role of dark energy is
the cosmological constant $\Lambda$ allowed by Einstein equations,
its smallness $\Lambda_{observed}\sim10^{-122}M_{p}^{2}$ is very
problematic as it requires an extreme fine-tuning. Besides, recent
observations indicate that a time dependent equation of state is
possible. In this case, the simplest approach is to consider a
scalar field to play the role of dark energy. There are many types
of such models and an extended literature, we refer only a few
 works in four of these approaches namely quintessence \cite{Peebles2}, k-essence
\cite{Chiba}, tachyons \cite{Sen} and quintom \cite{quintom}.

Moreover, the SNIa data admits an equation of
state $\omega_{de}<-1$, which is attributed to the so-called phantom dark
energy \cite{Caldwell}. This has the striking feature that
its density grows without limit with the expansion of the Universe. Usually, this
behavior leads to a violation of the weak energy condition and then, to the
so-called Big Rip in a finite time \cite{Bigrip}. This is true if
dark energy satisfies a conservation equation which corresponds to a
non-interacting fluid. Now, if dark energy interacts, for example with dark matter, then
the energy conservation equation is modified, and it is possible to circumvent
the blowing of the dark energy density. Further, there are
proposals that can encode the crossing $\omega_{de}=-1$ (dubbed phantom
divide) without violating the weak energy condition \cite{Aposto}. This behavior can be
exhibited in dark energy-dark matter unified models. An example of such a
model is the Chaplygin gas \cite{Kame} and its generalization
\cite{Bento}. The Chaplygin gas has a connection with string theory and can be
obtained from the light-cone parametrization of the Nambu-Goto action,
associated to a $D$-brane \cite{Kame2}. In this model, a single
self-interacting scalar field is responsible for both dark energy and dark
matter, giving also the observed accelerated expansion. In a recent work \cite{Bertolaminew},
it has been claimed that the galaxy cluster Abell A586 exhibits evidence that dark energy and
dark matter are interacting. The authors trace back the coupling to a departure of the
equilibrium settled by the virial theorem and show that it could be explained by the generalized Chaplygin
gas (GCG) with dark energy equation of state $\omega_{de}=-1$.

In the light of these last results and the considerations mentioned
above, in this paper we will consider the generalized Chaplygin gas as a
source of dark matter and dark energy which are interacting assuming a time dependent
dark energy equation of state with the property to encode the phantom-like behavior.
For simplicity, we will not consider the baryonic or other
kind of matter. A particular and simple ansatz on the
separation of the dark energy and dark matter densities is assumed.
 We find a phantom-like equation of state for
dark energy that crosses the phantom divide $\omega_{de}=-1$ in recent epoch. Its
associated density smoothly decays with the expansion of the
Universe (thus, phantom-like). It is also shown that the normalized
densities $\Omega_{i}$ are in the past in agreement with the
$\Lambda$CDM model, but in the future they enter in a phase of
mutual equilibrium, where their final values are practically the
same as today. This indicate a scaling solution, similar to
that obtained in \cite{Amen}, where it was assumed a stationary dark
energy which could solve the coincidence problem. Also, we find a coupling
that it is in agreement with one would expect in relation with the coincidence
problem.

We organized the present paper as follows: In section 2, we briefly
revisit the unified generalized Chaplygin gas. In Section 3, we construct
the model by splitting the total density into dark energy and dark matter
components, by means of our assumption. In section 4, we find a dark
energy equation of state and we analyze its general behavior. We
obtain also the involved densities explicitly in function of the
scale factor, and show that the model has a scaling solution. At the
end of this section, we plotted the relevant quantities for best
fitted values of the parameters. Finally, we present the conclusions
in section 5.

\section{The generalized Chaplygin gas}

\ In the GCG model the pressure $p$ of the fluid has the
following form \cite{Bento}%
\begin{equation}
p=-\frac{A}{\rho^{\alpha}}, \label{presseff}%
\end{equation}
where $\rho$ is the total density and $A>0$ and $\alpha\geq0$ are parameters. For
$\alpha=1$ we recover the standard Chaplygin gas and $\alpha=0$ corresponds to
the $\Lambda$CDM model. Following the definition $p\equiv\omega\rho$, we read
from (\ref{presseff}) the GCG equation of state (EoS)%
\begin{equation}
\omega=-A_{s}\left(  \frac{\rho_{0}}{\rho}\right)  ^{1+\alpha}, \label{eoseff}%
\end{equation}
where $A_{s}\equiv A/\rho_{0}^{1+\alpha}$ and $\rho_{0}$ is the total density
today, given by $\rho_{0}\equiv(3/8\pi G)H_{0}^{2}$ in terms of the Hubble
constant $H_{0}$ in a flat universe. In \cite{Bertolami} the ranges of the
values of $A_{s}$ and $\alpha$ have been analyzed, and it is argued that the
observations favor $0<A_{s}<1$ with $\alpha>1$, although $0<A_{s}<1$ with
$0<\alpha\leq1$ is still possible. In this work, we will apply our results in
both regions. The density satisfies the conservation equation%
\begin{equation}
d\rho+3\frac{da}{a}(\rho+p)=0. \label{chapcons}%
\end{equation}
Using (\ref{presseff}) in this last equation, we find the density in terms of
the scale factor%
\begin{equation}
\rho=\left[  A_{s}+\frac{1-A_{s}}{a^{3(1+\alpha)}}\right]  ^{\frac{1}%
{1+\alpha}}\rho_{0}, \label{denseff}%
\end{equation}
where we normalized the scale factor as $a_{0}=1$ today. The GCG EoS
(\ref{eoseff}) in terms of the scale factor is%
\begin{equation}
\omega=-\frac{A_{s}a^{3(1+\alpha)}}{1-A_{s}+A_{s}a^{3(1+\alpha)}}. 
\label{eoseff2}%
\end{equation}
It is easy to see that with the restriction $A_{s}>0$, this EoS is
constrained to the interval $-1\leq\omega\leq0$. Therefore, the
model is in general of quintessence and excludes the phantom region
$\omega<-1$, with a phase of acceleration for $\omega<-1/3$.
Furthermore, it allows us to interpret the fluid as cold matter
$\omega\simeq0$ for $a\rightarrow0$, and as dark energy
$\omega\simeq-1$ for $a\rightarrow\infty$ and it can be considered
as an unified model of matter and dark energy. If we assume in the
simplest case a EoS for dark energy $\omega_{de}=-1$, the resulting
model gives slight deviations of the $\Lambda$CDM model
\cite{Bento}. However, due to the fact that the model is a mixture
of dark matter and dark energy, we cannot exclude a priori the
possibility that the dark energy EoS $\omega_{de}$, has a phantom
phase. As the observations tend to support, it is possible that the
EoS for dark energy evolves in time and eventually crosses the
boundary $\omega_{de}=-1$, at a recent epoch. In fact, as we will
see below, that this is the case in the approach considered in this work.

\section{The Model}

Let us consider the Friedmann equation,%
\begin{equation}
H^{2}=\frac{8\pi G}{3}\rho,\label{fried}%
\end{equation}
where $H\equiv\dot{a}/a$ is the Hubble parameter and $\rho$ is the total density.
We assume that this density is decomposed as $\rho=\rho_{de}+\rho_{dm}$, where $\rho_{de}$ and
$\rho_{dm}$ are the densities of dark energy and dark matter respectively. Thus for cold dark matter, the conservation equation
(\ref{chapcons}) becomes
\begin{equation}
d\rho_{de}+3\frac{da}{a}(\rho_{de}+p_{de})=-\left(  d\rho_{dm}+3\frac{da}%
{a}\rho_{dm}\right)  .\label{interacting}%
\end{equation}
The problem now is to have a relationship between dark energy and
dark matter in order to solve (\ref{interacting}). To do it, in this
work we will assume a particular and simple ansatz which gives results in
agreement with the observations, as follows,
\begin{equation}
\rho_{de}^{2}=\lambda\rho_{dm},\label{fundamen}%
\end{equation}
where $\lambda$ is a constant. We will show that this is consistent
with the usual behavior of dark energy, suppressed at early times,
and then increasing and triggering the acceleration at late times.
We should note that equation (\ref{interacting}) has also been used in connection
with the generalized Chaplygin gas in holographic models \cite{Setare}.  In
contrast with our proposal (\ref{fundamen}), in these kind of models a holographic
dark energy density is assumed to be able to relate it with the
dark matter content.
Under ansatz (\ref{fundamen}), we find the normalized densities%
\begin{equation}
\Omega_{de} \equiv\frac{\rho_{de}}{\rho}=\frac{\lambda}{\lambda+\rho_{de}}
\qquad {\rm and}\qquad
\Omega_{dm}  \equiv\frac{\rho_{dm}}{\rho}=\frac{\rho_{de}}{\lambda+\rho_{de}},\label{normdens}
\end{equation}
which clearly satisfy the flat universe constraint $\Omega_{dm}+\Omega_{de}=1$.
If we impose the restriction $\rho_{de}\geq0$, then we have from
(\ref{normdens})%
\begin{equation}
\rho_{de}=\frac{\lambda}{2}\left[  \sqrt{1+\frac{4\rho}{\lambda}}-1\right]  .
\label{branedens}%
\end{equation}

\section{Dark energy equation of state}

Under the assumption of cold dark matter, we can identify the
pressures $p=p_{de}\equiv$ $\omega_{de}\rho_{de}$. Therefore,%
\begin{equation}
\omega\rho=\omega_{de}\rho_{de}, \label{equalpress}%
\end{equation}
and the dark energy EoS is,%
\begin{equation}
\omega_{de}=\omega\frac{\rho}{\rho_{de}}=\frac{\omega}{\Omega_{de}}.
\label{equalpress2}%
\end{equation}
After using (\ref{eoseff}), (\ref{normdens}) and (\ref{branedens}) in the last
equation, we obtain%
\begin{equation}
\omega_{de}=-\frac{A_{s}}{2}\left[  1+\sqrt{1+\frac{4\rho}{\lambda}}\right]
\left(  \frac{\rho_{0}}{\rho}\right)  ^{1+\alpha}. \label{eosde}%
\end{equation}
Substituting (\ref{denseff}) in this equation, we get the dark energy EoS
\begin{equation}
\omega_{de}=-\frac{A_{s}a^{3(1+\alpha)}}{2X}\left[  1+\sqrt{1+\frac{4\rho
_{0}X^{\frac{1}{1+\alpha}}}{\lambda a^{3}}}\right]  , \label{eosde2}%
\end{equation}
where
\begin{equation}
X\equiv1+A_{s}\left[a^{3(1+\alpha)}-1\right]>0.\label{X}
\end{equation}
We can see from (\ref{eosde}) that $\omega_{de}$ decreases as $a$
increases. Moreover, for enough large values of $a$, we find in
general that $\omega_{de}<-1$. If we consider, the today
value $a=1$, we get
\begin{equation}
\omega_{de0}=-\frac{A_{s}}{2}\left[  1+\sqrt{1+\frac{4\rho_{0}}{\lambda}%
}\right]  , \label{eosdetoday}%
\end{equation}
which, by considering the most probable values of $A_s$ given in
\cite{Bertolami}, we show below that it fulfills as well the phantom
divide condition $\omega_{de0}<-1$. Notice that once the boundary
$\omega_{de}=-1$ is crossed, dark energy behaves as phantom-like for
all the rest of the evolution of the Universe and it never returns
to the quintessence region $\omega_{de}>-1$. Moreover, in the limit
$a\sim0$, the dark energy EoS consistently approaches
$\omega_{de}\sim0$.

\subsection{Scaling Solution}

Scaling solutions are interesting as far as they could solve the coincidence
problem \cite{Amen}. In the scaling regime, the ratio $\rho_{dm}/\rho_{de}$ is
a non-zero constant. Thus, dark energy and dark matter remain
of the same order. We will see that our model has an asymptotic scaling region
for $a$ large. In order to see it, we compute the dark energy density by means
of (\ref{denseff}) and (\ref{branedens})%
\begin{equation}
\rho_{de}=\frac{\lambda}{2}\left[  \sqrt{1+\frac{4\rho_{0}X^{\frac{1}%
{1+\alpha}}}{\lambda a^{3}}}-1\right]  . \label{dedens}%
\end{equation}
Thus, taking into account (\ref{fundamen}) we get
\begin{equation}
\rho_{dm}=\frac{\lambda}{4}\left[  \sqrt{1+\frac{4\rho_{0}X^{\frac{1}{1+\alpha
}}}{\lambda a^{3}}}-1\right]  ^{2}. \label{dmdens}%
\end{equation}
These are solutions of equation (\ref{interacting}), taking into account
(\ref{eosde2}). Indeed, a simple calculation reproduces the GCG density (\ref{denseff}).
Substituting (\ref{denseff}) into (\ref{normdens}), we get the dark energy
normalized density in terms of the scale factor%
\begin{equation}
\Omega_{de}=\frac{2}{\sqrt{1+\frac{4\rho_{0}X^{\frac{1}{1+\alpha}}}{\lambda
a^{3}}}+1}, \label{normdens1}%
\end{equation}
as well as the dark matter normalized density
\begin{equation}
\Omega_{dm}=\frac{\sqrt{1+\frac{4\rho_{0}X^{\frac{1}{1+\alpha}}}{\lambda a^{3}%
}}-1}{\sqrt{1+\frac{4\rho_{0}X^{\frac{1}{1+\alpha}}}{\lambda a^{3}}}+1}.
\label{normdens2}%
\end{equation}
Then, the ratio is obtained %
\begin{equation}
\frac{\rho_{dm}}{\rho_{de}}=\frac{1}{2}\left[  \sqrt{1+\frac{4\rho_{0}%
X^{\frac{1}{1+\alpha}}}{\lambda a^{3}}}-1\right]  , \label{ratio}%
\end{equation}
which clearly differs from the non scaling models $\Lambda$CDM and GCG, with $\rho_{dm}/\rho_{de}\propto
a^{-3}$, and $\rho_{dm}/\rho_{de}\propto a^{-3(1+\alpha)}$ (see \cite{Bento}) respectively. For both models $\omega_{de}=-1$.

In our case, we find that this ratio is today
\begin{equation}
\frac{\rho_{dm0}}{\rho_{de0}}=\frac{1}{2}\left[  \sqrt{1+\frac{4\rho_{0}%
}{\lambda}}-1\right]  , \label{ratiotoday}%
\end{equation}
and for large values of $a$, we have a scaling solution
\begin{equation}
\frac{\rho_{dm}}{\rho_{de}}\simeq\frac{1}{2}\left[  \sqrt{1+\frac{4\rho
_{0}A_{s}^{\frac{1}{1+\alpha}}}{\lambda}}-1\right]  . \label{ratiofuture}%
\end{equation}
We can see that (\ref{ratiotoday}) and (\ref{ratiofuture}) are
practically the same for the allowed values of $A_{s}\sim 1$.
Therefore, dark energy and dark matter are of the same order today
and in the future, which alleviates the coincidence problem.
Further, the usual \textquotedblleft small\textquotedblright\ time
interval that takes the transition between dark matter and dark
energy (present for instance, in the $\Lambda$CDM model), in our
case stretches to infinity, in such a way that the dark energy and
the dark matter densities tend asymptotically to constant values
\begin{equation}
\rho_{de}\simeq\frac{\lambda}{2}\left[  \sqrt{1+\frac{4\rho_{0}A_s^{\frac{1}%
{1+\alpha}}}{\lambda}}-1\right]  ,\quad
\rho_{dm}\simeq\frac{\lambda}{4}\left[  \sqrt{1+\frac{4\rho_{0}A_s^{\frac{1}{1+\alpha
}}}{\lambda }}-1\right]  ^{2}. \label{rhos}
\end{equation}

Let us now consider the coupled equation for the interacting model
(\ref{interacting}) in the scaling regime, when
$\rho_{dm}$ and $\rho_{de}$ are practically constant. Then, $d\rho_{dm}\simeq0$
and $d\rho_{de}\simeq0$, therefore $p_{de}\simeq-(\rho_{dm}+\rho_{de})$. Thus, considering the EoS
$p_{de}=\omega_{de}\rho_{de}$, we get in general for this model
\begin{equation}
\frac{\rho_{dm}}{\rho_{de}}\simeq -(1+\omega_{de}).
\label{ratio2}%
\end{equation}
Hence,
\begin{equation}
\omega_{de}\simeq-(1+\frac{\rho_{dm}}{\rho_{de}})<-1.\label{26}
\end{equation}
Thus, in the scaling region dark energy EoS is phantom-like.
Because the ratio $\rho_{dm}/\rho_{de}$ in the future is practically
the same as today, we can put
$\rho_{dm0}/\rho_{de0}=\Omega_{dm0}/\Omega_{de0}\simeq0.3/0.7\simeq0.43$
in (\ref{26}). Then we get the lower bound $\omega_{de}\simeq-1.43$.
In this limit, the total density of
the GCG model (\ref{denseff}) is $\rho$ $\simeq
A_{s}^{\frac{1}{1+\alpha}}\rho_{0}$, and the Universe enters in the
de-Sitter phase $a(t)\simeq\exp(H_{f}t)$, where
$H_{f}^{2}\simeq(8\pi GA_{s}^{\frac {1}{1+\alpha}}\rho_{0})/3$.
Therefore, the Universe expands accelerated forever.
Now, we can analyze the coupling between dark energy and dark matter in more detail. We will follow
the definition of the coupling given in \cite{Bertolaminew}
\begin{equation}
\dot{\rho}_{dm}+3H\rho_{dm} =\zeta H\rho_{dm},\\
\dot{\rho}_{de}+3H\left(1+\omega_{de}\right)\rho_{de} =-\zeta
H\rho_{dm},\label{coupling}\\
\end{equation}where $\zeta$ is the coupling. Making the change $\dot{\rho}_{dm}=\dot{a}
d\rho_{dm}/da,\dot{\rho}_{de}=\dot{a}d\rho_{de}/da$ and using (\ref{fundamen})
we find
\begin{equation}
\zeta=-3\frac{(1+2 \omega_{de})}{1+2r},\label{coupling2}
\end{equation}
where we introduced the ratio $r\equiv\rho_{dm}/\rho_{de}$. We note that both fluids
always interact ($\zeta\neq0$), only when $\omega_{de}$ passes through $-1/2$ they uncouple.
Also notice that the coupling evolves from a negative value through
a positive value, once the condition $\omega_{de}<-1/2$ is satisfied.
Eventually when the approximation (\ref{ratio2}) is valid, this coupling
approaches to a non vanishing constant $\zeta\lesssim3$ for large values of $a$. This
is consistent with the fact that the fluid enters in the scaling regime. It is interesting to compare our model
with the work in \cite{Bertolaminew} where $\omega_{de}$ is considered constant. In that case
the coupling $\zeta'$ (we denote with primes the quantities in that work) results in
\begin{equation}
\zeta'=-\frac{(\eta+3 \omega'_{de})}{1+r'},\label{coupling2}
\end{equation}
which for the GCG model with $\omega'_{de}=-1$, one has $\eta=3(1+\alpha)$ and $r'=\rho'_{dm}/\rho'_{de}=\Omega_{dm0}/\Omega_{de0}a^{-\eta}$. Now, using (\ref{coupling}) for our case and a similar set of equations for the model
in \cite{Bertolaminew} we find the relation between $\zeta$ and $\zeta'$
\begin{equation}
\zeta=\frac{\rho'_{dm}}{\rho_{dm}}\frac{d\rho_{de}}{d\rho'_{de}}\zeta'-3\frac{\rho_{de}}{\rho_{dm}}(1+\omega_{de}),\label{relation}
\end{equation}
which shows that these couplings are different. Note, however, that both couplings coincide
if $\omega_{de}=-1$ and the densities with primes and those without primes coincide as well.

\subsection{Numerical Results}

For practical purposes and in order to be specific, we shall take numerical values
for the parameters involved in the model. Taking the present values for the
fractional densities as $\Omega_{de0}=0.7$ and $\Omega_{dm0}=0.3$, we obtain
from (\ref{normdens}) $\lambda=(\Omega_{de0}^{2}/\Omega
_{dm0})\rho_{0}\simeq1.63\rho_{0}$. The dark energy EoS (\ref{eosdetoday}) is today approximately $\omega
_{de0}\simeq-1.43\,A_{s}$, independent of $\alpha$, and is phantom-like for
$1>A_{s}\gtrsim 0.7$. This falls quite well in the region of confidence given
by the constraints on the observations for a flat Universe \cite{Bertolami}.
We will take the best fitted values in the two parametrizations:
$\alpha=.999$, $A_{s}=0.79$ for the range $0\leq\alpha\leq1$, and
$\alpha=3.75$, $A_{s}=0.936$ for the range $\alpha>1$ \cite{Bertolami}. In
Figure 1, we show the phantom-like behavior of $\omega_{de}$. We see that it
crosses the phantom divide at $a\simeq0.9$ for $A_{s}=0.79$ and has a value
$\omega_{de}\simeq-1.13$ today. In the case $A_{s}=0.936$, it crosses at
$a\simeq0.95$ and has a value $\omega_{de}\simeq-1.34$ today. In Figure 2, we
plot the quotient of the dark matter over dark energy densities for the same values as in Fig. 1.
We see that it approaches to a non vanishing constant limit as the Universe expands, and
it remains constant in this region. Further, from (\ref{ratiofuture}%
) we have in this region, $\rho_{de}\simeq2.55\,\rho_{dm}$ for $A_{s}=0.79$ and
$\rho_{de}\simeq2.35\,\rho_{dm}$ for $A_{s}=0.936$. In Figure 3, we plot
the coupling for the same choice of the parameters. As we can see, the coupling is
constrained to $\zeta<3$. Finally, in Figure 4 we
plot $\Omega_{dm}$ and $\Omega_{de}$. In the past they
behave as in the $\Lambda$CDM model, but in the future for $A_{s}=0.79$ they
approach to the limits $\Omega_{dm}\simeq0.282$ and $\Omega_{de}\simeq0.718$
whereas for $A_{s}=0.936$ they approach to $\Omega_{dm}\simeq0.298$ and
$\Omega_{de}\simeq0.702$. These are almost the same values as the ones
at present.

\begin{figure}[t]
\centerline{\includegraphics[height=8cm, width=12cm]{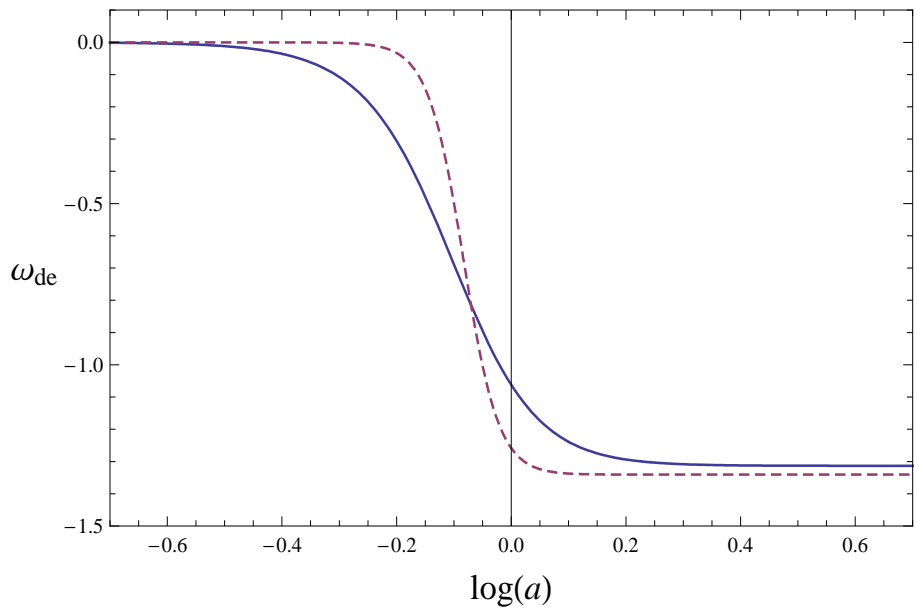}}
\caption{Dark energy EoS vs. log$a$. The solid line corresponds to $A_{s}=0.79$ and
$\alpha=0.999$. In this case dark energy EoS has a value $\omega_{de0}
\simeq-1.13$ today. The dashed line corresponds to $A_{s}=0.936$ and
$\alpha=3.75$ and $\omega_{de0}\simeq-1.34$ today. In both cases the phantom
divide is crossed recently.}
\end{figure}
\begin{figure}[t]
\centerline{\includegraphics[height=8cm, width=12cm]{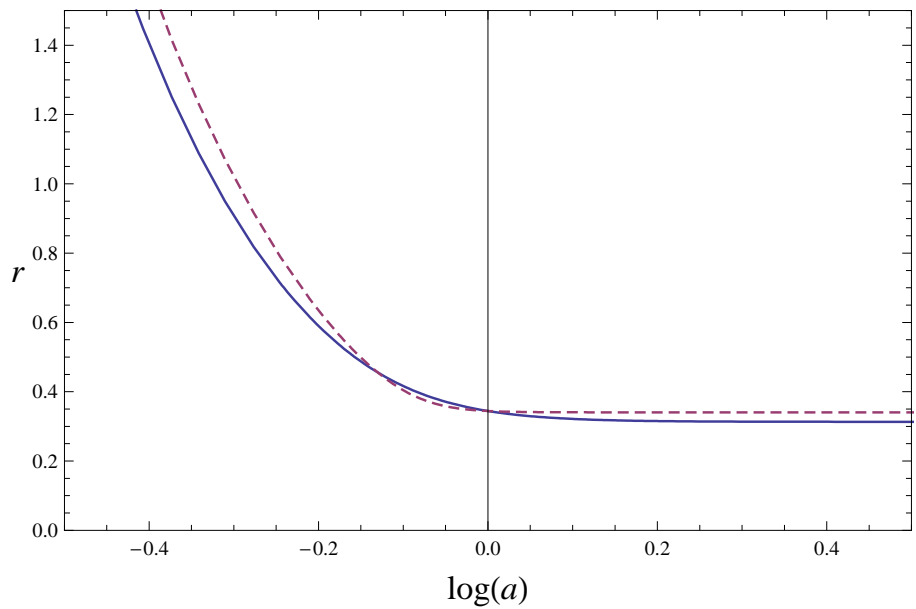}}
\caption{Dark matter density over dark energy density $r$ vs. log$a$ . The solid line corresponds to $A_{s}=0.79$ and
$\alpha=0.999$. The dashed line corresponds to $A_{s}=0.936$ and
$\alpha=3.75$.}
\end{figure}

\begin{figure}[t]
\centerline{\includegraphics[height=8cm, width=12cm]{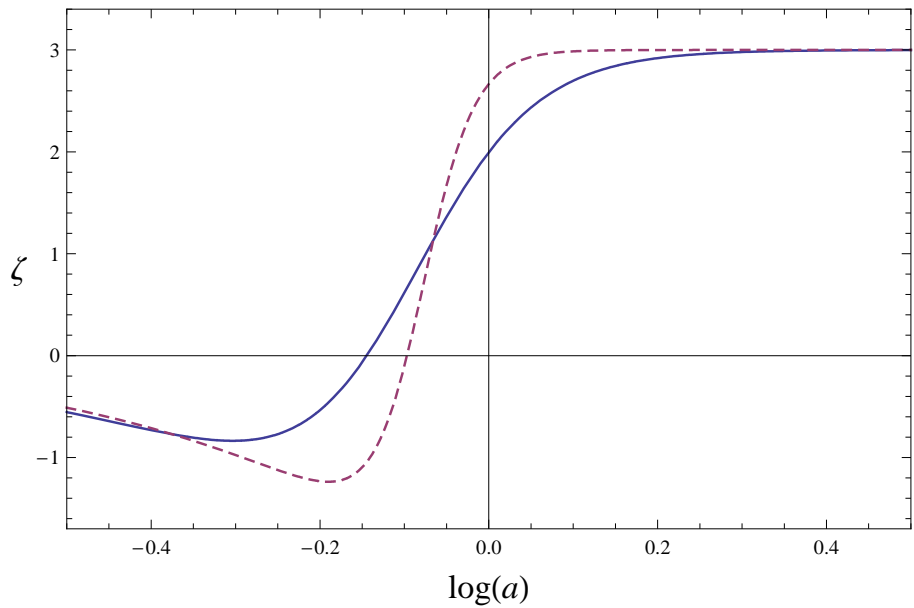}}
\caption{Behavior of the coupling $\zeta$ vs. log$a$. The solid line corresponds to $\alpha=0.999$ and $A_{s}=0.79$ and the dashed line
corresponds to $A_{s}=0.936$ and
$\alpha=3.75$.}
\end{figure}
\begin{figure}[t]
\centerline{\includegraphics[height=8cm, width=12cm]{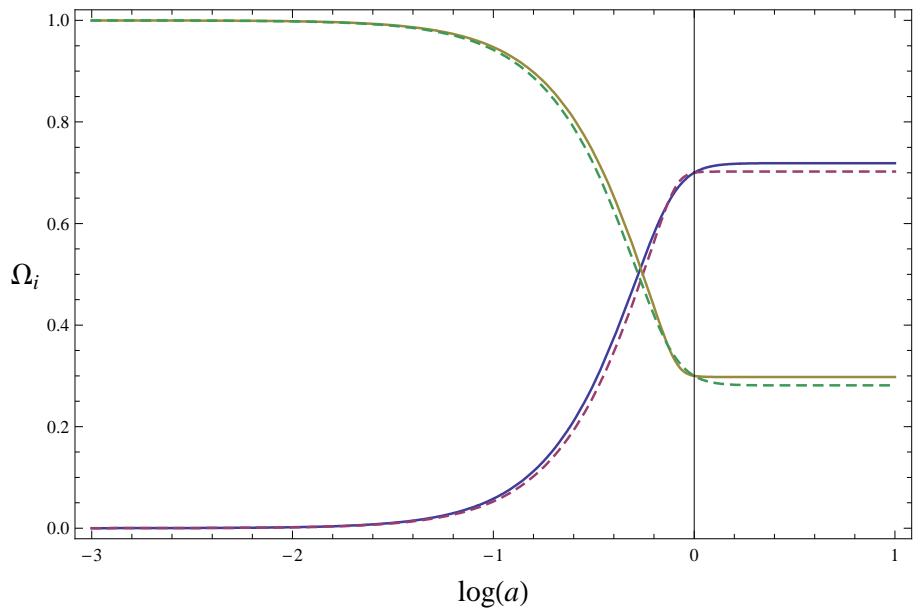}}
\caption{Normalized dark matter $\Omega_{dm}$ (decaying curves)
and dark energy $\Omega_{de}$ (growing curves) vs. log$a$. The solid lines correspond to $A_{s}=0.79$ and
$\alpha=0.999$. The dashed lines correspond to $A_{s}=0.936$ and
$\alpha=3.75$. We see that they behave like the $\Lambda$CDM
model in the past.}
\end{figure}

\section{Conclusions}

In this work we have studied the generalized Chaplygin gas as an
unified self-interacting fluid. By assuming a simple relation
between dark matter and dark energy densities (\ref{fundamen}) we
obtain a dark energy EoS which exhibits a phantom-like behavior. In
the regime of the most probable values of $A_{s}$ and $\alpha$
\cite{Bertolami}, this identification leads to a dark energy EoS
that crosses the phantom regime $\omega_{de}=-1$ in the recent time.
In fact, the numerical values $\omega_{de0}\simeq-1.13$ (for
$A_{s}=0.79$) and $\omega_{de0}\simeq-1.34$ (for $A_{s}=0.936$) are
consistent with the observations. Furthermore, the general behavior
of the dark energy EoS in our model, is in complete agreement with
the best fit in \cite{Alam} for $\Omega_{dm0}=0.3$. Once the dark
energy EoS crosses the phantom barrier, it never returns to
quintessence, giving a phantom-like sector for all the future
evolution of the Universe. The densities smoothly decay for all
values of the scale factor, thus avoiding a Big Rip. We find also,
that in contrast with the $\Lambda$CDM and GCG with $\omega_{de}=-1$ models \cite{Bertolaminew},
our model exhibits a scaling solution when $a$ is
large, as can be seen  from the numerical results. This
regime is achieved once the dark energy EoS has crossed the phantom
divide, and remains in this way for all the rest of the evolution.
We show also that the coupling between dark energy and dark matter
tends to a positive constant in this region and the Universe tends to be
de-Sitter. This type of solution could provide a clue to solve the
coincidence problem \cite{Amen}. We also find that the present model with time varying
EoS and with the assumptions made, could be considered at least as a phenomenological model
which could encode the interaction conjectured in \cite{Bertolaminew}, inferred from the Abell
cluster A586. The relation between both interacting models is
given through (\ref{relation}). It will be of interest to find scalar
fields and their corresponding potentials that could reproduce our
results. This will be studied in a future work.

\ack
This work was supported in part by CONACyT Grants Nos. 45713-F, 51306,
VIEP-BUAP grant No. 30/exc/07,  PROMEP UGTO-CA-3 and IAC. G. G. J thanks CONACyT
for a posdoctoral grant under the program Apoyos Integrales para la
Formaci\'{o}n de Doctores en Ciencias 2006.


\section*{References}
\begin{thebibliography}{99}                                                                                               %


\bibitem {Riess}A. G. Riess et al, Astron. J. \textbf{116}, 1009 (1998); S.
Perlmutter \textit{et al.}, Astrophys. J. \textbf{517}, 565 (1999).

\bibitem {Peebles}P. J. Peebles and B. Ratra, Rev. Mod. Phys. \textbf{75}, 559
(2003); E. J. Copeland, M. Sami and S. Tsujikawa, Int. J. Mod. Phys. D
\textbf{15}, 1753 (2006).

\bibitem {Peebles2}P. J. Peebles and B. Ratra, Astrophys. J. \textbf{325}, L17
(1988); B. Ratra and P. J. E. Peebles, Phys. Rev. D \textbf{37}, 3406 (1988);
L. A. Urena-Lopez and T. Matos, Phys. Rev. D \textbf{62}, 081302 (2000).

\bibitem {Chiba}T. Chiba, T. Okabe and M. Yamaguchi, Phys. Rev. D \textbf{62},
023511 (2000); M. Malquarti, E. J. Copeland and A. R. Liddle, Phys. Rev. D
\textbf{68 }, 023512 (2003).

\bibitem {Sen}A. Sen, JHEP \textbf{0204}, 048 (2002); A. Sen, Int. J. Mod.
Phys. A \textbf{18}, 4869 (2003); G. Gibbons, Phys. Lett. B \textbf{537}, 1
(2002); T. Padmanabhan and T. R. Choudhury, Phys. Rev. D \textbf{66}, 081301 (2002).

\bibitem {quintom}B. Feng, X. Wang and X. Zhang, Phys. Lett. B \textbf{607},
35 (2005); Y. Cai, H. Li, Y. Piao and X. Zhang, Phys. Lett. B \textbf{646},
141 (2007); Y. Cai, M. Li, J. Lu, Y. Piao, T. Qiu and X. Zhang, Phys. Lett. B
\textbf{651}, 1 (2007); Y. Cai, T. Qiu, Y. Piao, M. Li and X. Zhang, JHEP
\textbf{0710}, 071 (2007); T. Qiu, Y. Cai and X. Zhang, arXiv:0710.0115
[gr-qc] (2007).

\bibitem {Caldwell}R. R. Caldwell, Phys. Lett. B \textbf{545}, 23 (2002); R.
R. Caldwell, M. Kamionkowski and N. N. Weinberg, Phys. Rev. Lett. \textbf{91},
071301 (2003); R. R. Caldwell and M. Doran, Phys. Rev. D \textbf{72}, 043527 (2005).

\bibitem {Bigrip}B. McInnes, JHEP \textbf{0208}, 029 (2002).

\bibitem {Aposto}P. S. Apostolopoulos and N. Tetradis, Phys. Rev. D
\textbf{74}, 064021 (2006);

\bibitem {Kame}A. Kamenshchik, U. Moschella and V. Pasquier,
Phys. Lett. B \textbf{487}, 7 (2000), Phys. Lett. B
\textbf{511}, 265 (2001); N. Bilic, G. B. Tupper and R. D. Viollier, Phys.
Lett. B \textbf{535}, 17 (2002).

\bibitem {Bento}M. C. Bento, O. Bertolami and A. A. Sen, Phys. Rev. D
\textbf{66}, 043507 (2002); M. C. Bento, O. Bertolami and A. A. Sen, Phys.
Rev. D \textbf{70}, 083519 (2004); R. A. Sussman, arXiv:0801.3324 [gr-qc] (2008).

\bibitem {Kame2}A. Kamenshchik, U. Moschella and V. Pasquier, Phys. Lett. B
\textbf{487}, 7 (2000); R. Jackiw, Lectures on Fluid Mechanics Springer
Verlag, Berlin, (2002); N. Bilic, G. B. Tupper and R. D. Viollier, J. Phys. A
\textbf{40}, 6877 (2007).

\bibitem {Bertolaminew}O. Bertolami, F. Gil Pedro and M. Le Delliou, Phys. Lett. B
\textbf{654}, 165 (2007); arXiv:0705.3118v2 [astro-ph] (2008).

\bibitem {Amen}L. Amendola and D. Tocchini-Valentini, Phys. Rev. D
\textbf{64}, 043509 (2001).

\bibitem {Bertolami}O. Bertolami, A. A. Sen, S. Sen and P. T. Silva, Mon. Not.
Roy. Astron. Soc. \textbf{353}, 329 (2004); M. C. Bento, O. Bertolami, N. M.
C. Santos and A. A. Sen, Phys. Rev. D \textbf{71}, 063501 (2005).

\bibitem {Setare}M. R. Setare, Phys. Lett. B \textbf{654}, 1 (2007); Eur. Phys. J. C \textbf{50}, 991 (2007).

\bibitem {Alam}U. Alam, V. Sahni, T. D. Saini and A. A. Starobinsky, Mon. Not.
Roy. Astron. Soc. \textbf{354}, 275 (2004).

\end{thebibliography}
\end{document}